# A heat-and-pull rig for fiber taper fabrication

Jonathan M. Ward, Danny G. O'Shea, Brian J. Shortt, Michael J. Morrissey, Kieran Deasy and Síle G. Nic Chormaic

*Department of Applied Physics and Instrumentation, Cork Institute of Technology, Bishopstown, Cork, and Tyndall National Institute, Prospect Row, Cork, Ireland*

We describe a reproducible method of fabricating adiabatic tapers with 3-4 $\mu$m diameter. The method is based on a heat-and-pull rig, whereby a $CO_2$ laser is continuously scanned across a length of fiber that is being pulled synchronously. Our system relies on a $CO_2$ mirror mounted on a geared stepper motor in order to scan the laser beam across the taper region. We show that this system offers a reliable alternative to more traditional rigs incorporating galvanometer scanners. We have routinely obtained transmission losses between 0.1 and 0.3 dB indicating the satisfactory production of adiabatic tapers. The operation of the rig is described in detail and an analysis on the produced tapers is provided. The flexibility of the rig is demonstrated by fabricating prolate dielectric microresonators using a microtapering technique. Such a rig is of interest to a range of fields that require tapered fiber fabrication such as microcavity-taper coupling, atom guiding along a tapered fiber, optical fiber sensing and the fabrication of fused biconical tapered couplers.

# I. INTRODUCTION

Techniques for evanescent coupling of light into microspherical cavities include prism couplers,[1] optical fiber half-block couplers,[2] end-polished fibers,[3] and fiber tapers.[4] Fiber tapers have proven to be the most attractive device for achieving near lossless coupling of light into microspherical cavities and exciting the fundamental resonant mode.[4,5] The overlap of the taper and microsphere evanescent fields defines the strength of this coupling.[6] The attainment of taper diameters typically in the range of 1 to 4 $\mu$m is critical to maximize this coupling. For a 2 $\mu$m diameter taper at a wavelength of 1550 nm, the fraction of power in the core is nearly 96%.[7] Recent discussions on the form of the evanescent field have shown the spatial extent of the radial component of the field to greatly increase for diameters less than 3 $\mu$m.[7,8] It has also been reported that efficient coupling of light into submillimeter sized silica microspheres is possible for taper diameters up to 4.5 $\mu$m.[9]

The four most widely exploited means of achieving micron-sized tapers are by flame,[5,10,11] $CO_2$ laser heating,[12] microfurnace[13,14] and, to a lesser extent, fusion splicer.[4] While subwavelength diameters have been shown to be achievable with the flame method, it presents significant technical challenges.[11] Firstly, the gas flow rate must be precisely regulated in order to maintain a suitable temperature. The purity of the gas supply becomes increasingly important for smaller taper diameters due to contamination concerns. Air currents in the vicinity of the flame also pose a problem, thereby limiting the option of scanning the flame across a length of fiber and causing areas of uneven heating. Despite being able to produce submicron tapers with the microfurnace method, the fibers cannot be controllably structured to produce different taper profiles.

As an alternative, $CO_2$ lasers present a largely stable and easily controllable method of heating a fiber. Air currents or other deleterious environmental effects bear no consequence on the power output or the ability to scan the beam across the fiber with a mirror scanner. It is possible to precisely control the length of fiber to be heated (i.e. the hot-zone), thereby yielding any desired taper profile.[15] The physical process of heating a fiber with a laser beam involves the fiber absorbing radiation and heating from the inside, whereas for a flame the process involves heating the surface of the fiber. There is an inverse square relationship between radius and heating for a $CO_2$ laser heat source, while for a flame heat source there is simply an inverse relationship between heating and radius. This ultimately places a stricter limit on the minimum taper diameter attainable for a given $CO_2$ laser power as compared with a flame heat source.[16] Previous reported attempts of producing tapers using the $CO_2$ laser technique have achieved a diameter of 4.6 $\mu$m with a $CO_2$ laser power of 13 W and FWHM spot size of 820 $\mu$m using a galvanometer mirror scanner.[12] In this paper we describe a reliable method of fabricating low loss 3-4 $\mu$m diameter tapers as well as the possibility of fabricating bottle resonators using a 25 W $CO_2$ laser.

In recent years, interest in the use of microspherical resonators in cavity QED experiments has increased.[17] The use of such microcavities in these experiments requires the possibility of tuning the resonance frequency of the microcavity to an atomic line. In principle, there are two main methods of achieving this: strain tuning and temperature tuning. Strain tuning can be used to sweep the resonant frequency through the cavity FSR whereas temperature tuning is limited to a fraction of the FSR.[18] Temperature tuning is unsuitable as a stand-alone method and the fabrication of microspheres suitable for the strain tuning apparatus is difficult. A new type of prolate microcavity that offers potentially greater flexibility in tuning the microcavity resonance frequency has recently been described theoretically in the literature.[19,20] Such cavities are termed *bottle resonators*. Strain tuning of these devices may tune the frequency over several FSR, while temperature tuning over a single FSR may only require a few tens of Kelvin. Another appealing feature compared with typical spherical microcavities is the stronger evanescent field at the bottle resonator surface due to the smaller resonator dimensions.

Our fabrication method relies on the use of a geared stepper motor to scan the laser beam rather than the more traditional galvanometer. Implementation of our rig is trivial, requiring only interconnection of the stepper motor with the controller circuit. In contrast, galvanometer scanners require PD (proportional-derivative) or PID (proportional-integral-derivative) control, which can be tedious to tune correctly. From a mechanical perspective, the stepper motor has a more robust design and better torque which negates any effects induced by the inertia of the mirror attached to the shaft of the motor and the overall scanning angle achievable is larger than that for a galvanometer scanner (typically limited to ±40°). Additionally, the stepper motor represents a significant cost saving since the cost of a scanning galvanometer system is typically in excess of US$1990,[21] while the cost of a Radionics hybrid stepper motor, gear box and controller circuit is approximately US$210.[22] By choosing the stepper motor scanner rather than a galvanometer scanner we saved 15 % on the total laser rig cost.

The technique described is not only of interest for applications involving micro-resonator coupling; areas requiring the fabrication of fused biconical tapered couplers will also find this inexpensive apparatus of use. This paper describes the procedure and requirements for pulling adiabatic fiber tapers[15,23] with typical losses in the range of 0.1 to 0.3 dB at 980 nm. We also report on successful attempts at fabricating bottle resonators to suitable dimensions using a microtapering technique.

## II. EXPERIMENT

A schematic of the laser scanning rig is shown in Fig. 1. Custom designed software automates the instrument control and data acquisition through serial and USB interfaces, thereby ensuring ease of parameter adjustment and reproducibility. A 12.5 cm focal length ZnSe lens focuses the laser beam onto the fiber with a FWHM spot size of approximately 500 $\mu$m, 5.5 cm in front of the scanning mirror. The stepper motor used in conjunction with a 500:1 gear box in half-step mode has a resolution of $1.8\times10^{-3}$ degrees per half-step, which corresponds to a hot-spot translation of around 1.7 $\mu$m per half-step on the fiber. The *hot-spot* refers to the point on the fiber which is being heated by the laser beam at any instant. A 1" gold mirror and holder are vertically mounted on the shaft of the gear box. The $CO_2$ laser beam is incident on the center of the mirror at an angle of 45° and is directed onto the fiber. A Thurlby Thander TG250 function generator connected to the stepper motor controller circuit cyclically scans the mirror through a sweep angle of approximately 10 degrees, thereby scanning the laser beam across the fiber. Alternatively, we can use customized limit switches to control the sweep angle of the mirror. Another function generator connected to the stepper motor controller circuit provides the clocking pulses, which dictate the speed at which the mirror scans the laser beam. The discrete nature of the stepper motor step-size is smeared out by the size of the beam at the focus. Two motorized translation stages (Standa Ltd., Lithuania) pull the ends of the fiber taper (SM980 single mode fiber from Fibercore) with a resolution of 1 $\mu$m. As the taper is being drawn, the transmission loss of a 980 nm diode laser through the fiber is monitored with a Thorlabs Si-photodiode (DET-series) connected to a digital storage oscilloscope.

The computer provides an analog voltage signal to the UC-2000 laser controller based on the power curve shown in Fig. 2. Although the power curve is an oversimplification,[16] we find it serves the application quite well. To determine the initial settings for our software, we manually increased the power of the laser while observing the incandescent light produced at the hot-spot. The laser power was recorded as a function of pull length. A fourth-order polynomial fit to this data provides the laser power curve. An exponential fit was also attempted but the slope of the curve increased too rapidly towards the end of the taper pull causing premature breaking of the fiber and limiting the minimum taper diameter to 4 $\mu$m. On the other hand, the polynomial power curve readily produced 3 and 4 $\mu$m diameter tapers.

Since the taper profile closely follows an exponential profile, we use this relationship to predict suitable scan lengths, *L*, and pull lengths, *z*, for any desired taper waist radius *r(z)*:

$$r(z) = r_0 e^{-z/L} \qquad (1)$$

where $r_o$ is the initial fiber radius before tapering.[15] There is some flexibility in choosing *z* and *L*. Scan lengths ranging from 5 to 15 mm, and pull lengths ranging from 12 to 40 mm were examined and all yielded low transmission losses and taper diameters of 3-4 $\mu$m. We found that Eq. (1) always gives an accurate prediction of the waist diameter. The pull speed is less flexible and a value of 80-110 $\mu$m per second provided the best results.

Several preparatory steps are essential in ensuring that the finished taper is of satisfactory quality. Firstly, the fiber is liberally cleaned with acetone to remove inorganic substances. It is imperative that the laser beam and fiber be horizontal to within a few tens of microns so that the focus of the beam covers the fiber precisely throughout the scan. The fiber must also be slightly taut before attempting this alignment. Failure to optimize the alignment before starting fabrication results in sagging and possible vibrations of the fiber, which tends to distort taper profiles leading to high transmission losses. The polarization of the laser beam is set vertical to the fiber so that the absorption coefficient is maximized.[16]

## III. RESULTS AND DISCUSSION

As a rule, tapers can only be produced with low losses if their profile follows the adiabaticity criterion.[23] This criterion requires the taper profile to be such that the change in taper angle is small enough to prevent light propagation being either coupled from the fundamental mode, $HE_{11}$, to higher order parasitic modes in the fiber or being lost as radiation. The criterion can be stated as $|dr/dz| \leq \rho(z)[\beta_1(z) - \beta_2(z)]/2\pi$, where $dr/dz$ defines the local taper angle, $\Omega(z)$, by the trigonometric expression $\Omega(z) = \tan^{-1}|dr/dz|$, $\rho(z)$ is the local core radius, $\beta_1(z)$ and $\beta_2(z)$ are the local propagation constants of the fundamental mode and the next closest mode respectively, and $z$ is along the fiber axis.[15,23] The transcendental equation for the propagation constants has been solved numerically as a function of local taper diameter. Fig. 3 shows a plot of this criterion and the core taper angle of a typical adiabatic taper as produced by the above-described method. The taper curve is based on an optimum exponential fit of the measurements in Fig. 4 and has the form as described by Eq. (1).

When the angle of the delineation curve is less than approximately ten times the taper angle, the light in the propagating mode will be lost to parasitic modes. While the fiber has only a single propagating mode in the untapered region, the taper itself contains multiple modes since the light is cladding-air guided. For a taper waist radius, $a$, of 1.5 $\mu$m, the core guidance parameter $V \equiv ka\sqrt{n_{core}^2 - n_{clad}^2} \approx 9.669 > 2.405$ at a wavelength of 980 nm, indicating the taper is multimoded.[8] The refractive index of the core, $n_{core}$, is actually the effective index of the fundamental mode determined from $\beta_1(z)/k$ where $k$ is the free space wavenumber, and the refractive index of the air cladding, $n_{clad}$, is 1. However, once the taper is adiabatic, light only propagates in the fundamental mode. A single-mode tapered fiber is only possible when the diameter is less than about 1 $\mu$m.

The delineation curve in Fig. 3 is read from right to left when considering a decreasing taper diameter. Starting from the untapered region on the right, the initial taper transition consists of a core where the light propagates and a cladding where the refractive index difference is $\Delta = (n_{core} - n_{clad})/n_{clad} \approx 7.8 \times 10^{-3}$. An approximation for the core guidance curve is given in [23], and is based on the assumption of an infinite cladding diameter. Gradually, the diameter and core guidance parameter decrease until the core effectively disappears and the propagating light becomes cladding-air guided. At the taper waist, there is a large index difference of about 0.42 between the cladding and surrounding air causing the core guidance parameter to gradually increase; this is shown as an increase in the angle of the delineation curve. For the cladding guidance curve, an approximation is also given in [23], and is based on the assumption that the core diameter is vanishing. Figure 4 shows a typical taper angle decreasing when approaching the taper waist. Both core and cladding contribute to distinct regions in the delineation curve. The taper clearly meets the adiabaticity criterion since the angle of the taper transitions is sufficiently below the delineation curve.

Fig. 5 shows power transmission loss for a taper with a 3 $\mu$m diameter as a function of pull length. The power remains constant except for minute oscillations that do not influence the final taper transmission. A final transmission loss of 0.20 dB was measured and is quite acceptable for applications involving microsphere coupling. For a set of 11 taper pulls, the loss was 0.3 ± 0.2 dB with some of the tapers having losses as low as 0.04 dB. Losses lower than 0.04 dB were not routinely observed, which is thought to be primarily due to the fact that the taper curve is not strictly a factor of 10 below the delineation curve for the whole taper profile. Using a shorter wavelength diode laser would proportionately shift the delineation curve to the left and this would increase the taper-delineation curve separation in Fig. 3, consequently giving lower losses.[23] Surface roughness on the taper is negligible as evidenced by the SEM image in the inset of Fig. 4.

The bottle resonator in Fig. 6 is produced with similar dimensions to that described in the theoretical work of Louyer *et al.*[19] and Sumetsky.[20] The resonator has a midsection diameter of 12 $\mu$m and length of approximately 300 $\mu$m. Shorter bottle resonators are limited by the size of the hot-spot. Kakarantzas *et al.*[24] describe a bottle resonator 160 $\mu$m long with a diameter of 16 $\mu$m. The fabrication begins with tapering of the fiber down to a waist diameter of 10-20 $\mu$m as described already. Each microtapered section on either side of the resonator is produced by pulling the taper at a speed of 10 $\mu$m per second for a length of about 0.2 mm while keeping the hot-spot stationary. The discrete rotation of the stepper motor shaft allows the scanning mirror to direct the beam onto exact locations where microtapering is to take place.

Subwavelength diameters of below 100 nm have been achieved with the microfurnace method.[14] A 20 W $CO_2$ laser heats a small Sapphire tube that surrounds the fiber rather than directly heating the fiber with the laser. Alignment of the beam focus with the taper is not as stringent as compared to the direct laser heating method. A limitation of this method is that microstructuring of the fiber profile cannot be achieved. Different taper shapes[15] may not be easily produced since the hot-zone cannot be precisely varied and is typically several millimeters in length. A very short hot-zone is vital for the microtapering step when producing bottle resonators. Furthermore, high losses of several dB have, until now, been observed with the microfurnace method because the taper transitions are non-adiabatic.[13] This is likely to be detrimental when coupling to the fundamental mode of a microresonator.

## IV. CONCLUSIONS

We have described a simple and reproducible method of fabricating 3-4 $\mu$m diameter tapers using commercial single-mode SM980 fiber. A stepper motor and gear box can be used to scan the laser beam across the fiber with sufficient precision as a simpler and inexpensive alternative to the more traditional galvanometer scanners reported for flame systems. Transmission losses of 0.3 ± 0.2 dB at a wavelength of 980 nm are within acceptable limits of other tapering rigs described in the literature.[25] We verify our optical microscope measurements of the taper diameter with SEM imaging. This rig is of interest for applications involving micro-resonator coupling and it can also be used to produce long period gratings[24] and fused biconical taper couplers.[25]

Apart from the simplicity and cost savings of such a laser-scanning rig, we have shown its flexibility for producing bottle resonators. This technique involves microtapering a fiber over a distance of about 0.2 mm and at a very slow speed. The dimensions of this new type of resonator can be precisely controlled. The discrete step-size of the stepper motor can be used to control the length of the resonator with a resolution of approximately 1.7 $\mu$m per half-step given the current rig configuration. The diameter of the bottle resonators can be as small as 3-4 $\mu$m. Installation and operation of our laser scanning rig requires a minimum amount of skill compared with most other tapering methods.

Smaller diameter tapers and shorter bottle resonators would, no doubt, be possible by first expanding the $CO_2$ beam before focusing it through the 12.5 cm lens, thereby yielding a higher power density by virtue of the smaller size of the laser beam at the focus. This will be the subject of future investigations. Even though the microfurnace tapering method can achieve any reasonable diameter required for microcavity coupling, the technique we have described is still required for bottle resonator fabrication. We intend to use these tapers to couple light into active microspheres made from doped phosphate and ZBNA glass.[26] We are also interested in studying the interactions between cold, rubidium atoms and the evanescent field at the taper region for atom manipulation and guiding purposes.[27]

## V. ACKNOWLEDGEMENTS

This work is supported by Science Foundation Ireland project number 02/IN1/128. We thank Dr. Anthony Grant of Cork Institute of Technology for his help with the SEM measurements. DOS acknowledges support from IRCSET through the Embark Initiative. KD is supported by Cork Institute of Technology and MM acknowledges support from the Council of Directors of the Institutes of Technology. The authors wish to thank one of the referees for drawing their attention to the microfurnace method described in [13, 14].

**Figure Captions**

Fig. 1.   Schematic of the taper fabrication rig. Dashed lines indicate control lines or data transfer.

Fig. 2.   Power curve for the $CO_2$ laser for a pull length of 22 mm.

Fig. 3.   Approximated length-scale delineation curve.

Fig. 4.   Taper profile for a 3 $\mu$m diameter taper. The pull length is 40 mm, scan length is 7 mm and the initial taper diameter is 125 $\mu$m. The inset shows an SEM image of a section of the taper with a diameter of 5.9 $\mu$m. The bar is 1.0 $\mu$m.

Fig. 5.   Transmission loss measurements for a 3 $\mu$m diameter taper with 0.20 dB loss.

Fig. 6.   Optical micrograph of a bottle resonator with a diameter of 12 $\mu$m and length of 300 $\mu$m. The bar is 10 $\mu$m.